\documentclass[conference]{IEEEtran}
\IEEEoverridecommandlockouts
\usepackage{moreverb,url}
\usepackage{enumitem}
\usepackage{cite}
\usepackage{amsmath,amssymb,amsfonts}
\usepackage{algorithmic}
\usepackage{graphicx}
\usepackage{textcomp}
\usepackage{xcolor}

\usepackage{textcomp} 
\usepackage{amssymb}  
\usepackage{bm}       
\usepackage{booktabs} 
\usepackage{dcolumn}  
\usepackage{multirow} 
\usepackage{graphicx} 
\usepackage[printonlyused]{acronym} 

\def\BibTeX{{\rm B\kern-.05em{\sc i\kern-.025em b}\kern-.08em
    T\kern-.1667em\lower.7ex\hbox{E}\kern-.125emX}}
\begin{document}

\title{Pre-trained Under Noise: A Framework for Robust Bone Fracture Detection in Medical Imaging\\
}
\author{
\IEEEauthorblockN{Robby Hoover}
\IEEEauthorblockA{\textit{School of Information Technology} \\
\textit{University of Cincinnati}\\
Cincinnati, Ohio \\
hooverrp@mail.uc.edu}
\and
\IEEEauthorblockN{Nelly Elsayed}
\IEEEauthorblockA{\textit{School of Information Technology} \\
\textit{University of Cincinnati}\\
Cincinnati, Ohio \\
elsayeny@ucmail.uc.edu}
\\  
\IEEEauthorblockN{Chengcheng Li}
\IEEEauthorblockA{\textit{School of Information Technology} \\
\textit{University of Cincinnati}\\
Cincinnati, Ohio \\
li2cc@ucmail.uc.edu}
\and
\IEEEauthorblockN{Zag ElSayed}
\IEEEauthorblockA{\textit{School of Information Technology} \\
\textit{University of Cincinnati}\\
Cincinnati, Ohio \\
elsayezs@ucmail.uc.edu}
}

\maketitle

\begin{abstract}
Medical Imagings are considered one of the crucial diagnostic tools for different bones-related diseases, especially bones fractures. This paper investigates the robustness of pre-trained deep learning models for classifying bone fractures in X-ray images and seeks to address global healthcare disparity through the lens of technology. Three deep learning models have been tested under varying simulated equipment quality conditions. ResNet50, VGG16 and EfficientNetv2 are the three pre-trained architectures which are compared. These models were used to perform bone fracture classification as images were progressively degraded using noise. This paper specifically empirically studies how the noise can affect the bone fractures detection and how the pre-trained models performance can be changes due to the noise that affect the quality of the X-ray images. This paper aims to help replicate real world challenges experienced by medical imaging technicians across the world. Thus, this paper establishes a methodological framework for assessing AI model degradation using transfer learning and controlled noise augmentation. The findings provide practical insight into how robust and generalizable different pre-trained deep learning powered computer vision models can be when used in different contexts.
\end{abstract}


\section{Introduction}
Disparity in healthcare outcomes remains one of the most pressing global challenges of our time. Access to quality medical diagnostics varies dramatically between nations, and even between social strata within nations. In this regard, medical imaging is at the center of enabling the provision of healthcare services, especially in the diagnosis of bone fractures through the use of X-ray technology. However, the quality of imaging equipment available in resource-limited settings often falls short of optimal standards, which can undermine diagnostic precision, as well as patient treatment quality \cite{destigter2021}.

Recent advances in the field of artificial intelligence, particularly with deep learning, offer promising solutions to bridge this gap in healthcare outcomes. Deep learning models have shown remarkable performance with medical image analysis, potentially providing consistent and accurate diagnostics regardless of geographic location or resource availability~\cite{lee2017deep,pellakur2023convolutional,razzak2017deep,abunajm2023deep,murtaza2020deep,desai2022transfer,suzuki2017overview}. However,  real-world implementation of these systems in resource-limited settings presents unique challenges, particularly in terms of the quality of available imaging equipment \cite{compton2018}.

This paper aims to investigate the robustness of deep learning models in medical imaging diagnostics under varying equipment quality conditions. Specifically, transfer learning has been empirically investigated on three different image classification models for bone fracture detection in various regions of the body, using a freely available dataset. These images go through several iterations in which the data is augmented with progressively more artificial noise. Various types of artificial noise are used, serving as analogs for common types of noise interference found in low-quality medical imaging equipment. 

Thus, this paper main research focus can be summarized at the following research question:

\emph{RQ: How does the performance of pre-trained deep learning image classification models degrade as their input is progressively more affected by different types of simulated medical imaging noise?}

To address this research question, first, the data augmentation task is performed using a variety of pre-processing techniques. These images are then used to fine-tune a series of CNN-based image classification models, which have been previously trained on other datasets, most notably ImageNet. Several iterations of the test dataset will be fed into the model with varying degrees of digitally simulated noise and the performance metrics will be compared over time as more noise is added to the test input. 

This paper also provides a systematic analysis of the different types of noise common in medical imaging. It establishes a methodology for evaluating AI model performance under varying equipment quality conditions, and offers practical insights into the feasibility of deploying advanced AI diagnostic tools in resource-limited settings.

\section{Background and Related Work}

\subsection{X-ray Medical Imaging}
X-rays are the most popular form of medical imaging technology in use today~\cite{mustapha2021comparative,wolbarst1999looking}. They serve as a cost effective method for identifying hard tissues and bone structures inside the body, and are most commonly used to identify bone fractures. When exposed to X-ray radiation, calcium in the bones will absorb the radiation and appear more white, while tissues will appear more black. Consequently, bone fractures will appear as dark areas within the otherwise white area created by the bone \cite{goyal2018,hoheisel2006review}.

However, obtaining an X-ray image has some unique challenges to it, as all forms of medical imaging do~\cite{ritman2006medical}. X-ray images may be corrupted by a variety of statistical noises, which can seriously deteriorate the quality of the images and can make diagnosing problems based on them a far more difficult task \cite{sun2018,huda2015x}.

Due to the physical processes involved in the capture of medical images, the resultant images are often quite noisy~\cite{goyal2018noise,gravel2004method}. Anything from electrical interference to distribution of light to quality of parts used can effect the quality and usability of the resultant image. Removal of noise from these images is typically seen as essential to enhancing the outcomes of human-reviewed medical images. A good denoising algorithm can enhance quality and even recover anatomical details that would have otherwise been hidden  \cite{rodrigues2008}. In general, noise is a significant factor in the processing and interpretation of medical images~\cite{sprawls2014optimizing}. 

\subsection{Deep Learning and Analyzing Medial Imaging}
Deep learning has been employed in different detection, classification, and prediction tasks in computer vision, where it has shown significant success in enhancing the system's performance~\cite{voulodimos2018deep,adewopo2023review,chai2021deep,affonso2017deep,adewopo2024smart,elsayed2019reduced,zhang2019medical,brownlee2019deep,adewopo2023baby,mikolajczyk2018data,Elsayed_ElSayed_Abdelgawad_2025,darlow2017fingerprint,shapira2019flowpic}. Deep Learning solutions for analyzing medical imaging for classification have been shown to have  performance comparable to that of humans~\cite{liu2019comparison}. Physicians are experiencing an all-time high in the number of complex readings they need to make \cite{kim2019}, especially in resource-limited environments \cite{patel2003}. Deep Learning solutions, especially those based in computer vision, have been proposed to create tools to help augment the duties of the physician by creating a collaborative medium that can decrease the total burden by relieving them of highly repetitive tasks~\cite{esteva2021deep,javaid2024computer,elyan2022computer,esteva2019guide}. 

However, solutions based in machine learning are not without challenges. Supervised algorithms, such as those most commonly used in deep learning computer vision tasks are trained on the assumption of a (mostly) clean dataset \cite{atla2011}. Previous work has been done to study the effects of noise on traditional machine learning methods, which have generally found that noise tends to cause model performance to decay as more noise is applied. However, this sort of work has not been applied to the field of deep learning hardly at all, and even less so for computer vision tasks. This previous work also inspected types of noise which are not directly applicable to the field of computer vision, which will require its own analysis of the types of noise unique to it. 

In this paper, we investigate three pre-trained and fine-tuned image classification architectures. The first architecture we are considering is the ResNet50 model. ResNet50 is a 50-layer convolutional network that pioneered the use of residual connections to combat the vanishing gradient problem and allow for much deeper networks \cite{he2016}. The VGG16 network is the second architecture under consideration. VGG16 is a 16-layer convolutional network which is generally renowned for its simplicity. It is constructed from a series of convolutional and pooling layers of increasing depth \cite{simonyan2014}. Finally, we also consider EfficientNetV2-S. EfficientNet is known for its distinctive architecture, which creatively balances its size and power, leading to quicker identifications in a shorter period of time \cite{tan2021}. Table~\ref{tab:model_comparison} shows a comparison between the three selected pre-trained models.

\begin{table}[h]
\centering
\caption{Comparison of Pre-trained Deep Learning Models based on Architecture Characteristics}
\begin{tabular}{|l|c|c|c|}

\hline
\textbf{Compare} & \textbf{ResNet50} & \textbf{VGG16} & \textbf{EfficientNetV2-S} \\
\hline
No. Layers & 50 & 16 & 173\\
\hline
Parameters & 25.6M & 138M & 21.5M \\
\hline
\end{tabular}
\vspace{0.5cm} 

\label{tab:model_comparison}
\end{table}

\section{Methodology}

\subsection{Dataset}
The dataset is prepared for use by applying a series of augmentations and pre-processing techniques to the images in the dataset. A series of image classification models are fine-tuned using this augmented data in order to create models which can take in an X-ray image as input, and output a predicted class, either containing a fracture, or not containing a fracture. Noise is progressively added to each model's evaluation input in order to determine how increasing amounts of noise affect the performance of each model.

For this experiment, the dataset used is the "Bone Fracture Multi-Region X-ray Data" dataset \cite{rodrigo2024}. It is freely available online with the Open Data Commons Attribution License. This dataset contains 10,580 radiographic images, of which 9246 are split for training, and the rest are split between testing and validation. This dataset comprises fractured and non-fractured X-ray images covering all anatomical body regions, including lower limb, upper limb, lumbar, hips, and knees. 

\subsection{Data Preprocessing}
Several techniques are used to augment the dataset to improve learning. First, images are resized to 180x180 pixels to maintain uniform size. This is done to ensure that all images have the same size, which is a requirement of many convolution-based image classification architectures \cite{hashemi2019}. Then, random horizontal flips and random affine transformations (scaling, shearing and slight translation) are applied. Rotation is not be specifically applied, as the dataset is already augmented with rotational variants. Images are normalized using ImageNet's standard parameters, as all three models were pre-trained on this dataset \cite{simon2016}.

\begin{figure}
    \centering
    \includegraphics[width=1\linewidth]{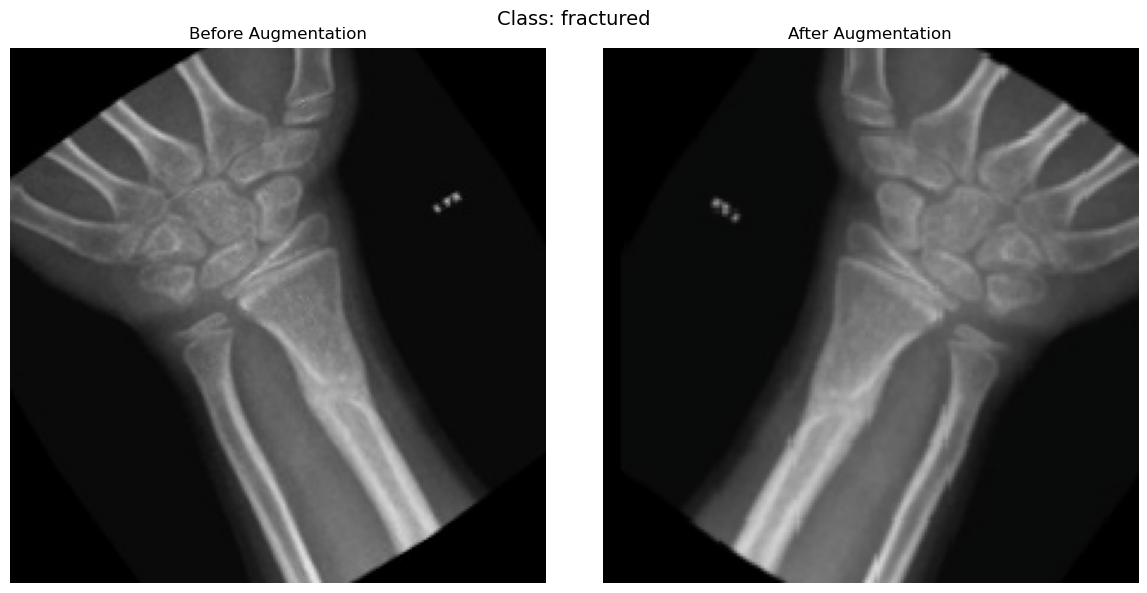}
    \caption{An image from the 'fractured' class before and after augmentations are applied.}
    \label{fig:enter-label}
\end{figure}

An 87\% : 8\% : 5\% ratio is the provided split between the training, validation and testing data subsets, respectively. The training set is used by the model as the ground-truth examples to learn from during the training phase. The validation set will be used to evaluate the model's performance during training, guide hyperparameter tuning, and determine when to stop training via early stopping. The testing set will be used as a completely independent dataset to evaluate the final model's performance on unseen data, providing an unbiased estimate of the model's real-world effectiveness. Figure\ref{fig:enter-label} demonstrates an example of an image before and after it is processed. 

\subsection{Transfer Learning and Fine Tuning}
To fine-tune the model, the layers and weights are initially frozen, in order to preserve the latent features found in the pre-trained weights. Then, parts of the model are progressively unfrozen until the model reaches an accuracy score higher than 95\%. This results in slightly different fine-tuning procedures for each model. 

The ResNet50 model requires the unfreezing of the third and fourth feature extraction layers, as well as unfreezing the entire classification layer. EfficientNet requires the last three layers of the feature extractor to be unfrozen and the classification layers to be unfrozen. The VGG-16 network only requires that the classification layer be unfrozen, without touching the feature extraction layers.  

As a baseline reference to compare the results of fine tuning against, we are planning to use a modern, simplified variant of the VGG network model. This "Simple CNN" network uses four 3x3 convolutional blocks, and utilizes the same design strategy of stacking convolutional transforms and nonlinear transforms. It also ends with a couple of fully connected classification layers. The only non-VGG modification that has been made to this network is to make full use of batch normalization and adding regularization to the classification layers. 

All models are trained with a learning rate of $lr = 0.001$ and trained for five epochs. All models will have their final classification layer replaced with a new layer to match the binary classification problem. All pre-trained weights used are obtained from the Torchvision computer vision library. 

\subsection{Digitally Simulated Noise}
Noise is any undesirable signal that is added into a desired signal at the time of that signal's acquisition from the machine which creates it. In medical imaging, noise can affect the intensity level of one pixel or many pixels and it typically results in a poorer image quality, which may make image classification harder for that given image \cite{kociolek2020}.

In a standard plain-film X-ray, the quantum noise (quantum mottle), structural noise, and electronic noises are the main sources of noise \cite{whitehead2025}. From the main sources of X-ray noise, we will be modeling quantum noise and electronic noise. In a comparison of various radiographic techniques, X-ray radiographs were found to be most prone to Poisson and Gaussian noise distributions \cite{goyal2018}, which correspond with the distributions of quantum and electronic noise. 

Structural noise will be ignored. Structural Noise has to do with the superposition of various anatomic structures, for example, how ribs line up with lungs during a radiograph \cite{hanson1981}. This kind of noise is fixed pattern noise rather than a random statistical process and, therefore, is not suitable for this study.  

Quantum noise, which is also sometimes known as quantum mottle, photon noise or shot noise, is the most significant source of noise in standard X-ray photography. It is a stochastic process due to the fluctuations in the number of photons which reach the detector from point to point. The number of photons which reach the detector follows a poison distribution, and therefore the resulting noise pattern also follows a Poisson distribution \cite{bell2025}.

Despite quantum noise being the generally prevailing type of noise that we see in X-ray imaging systems, we also frequently see noise generated by fluctuations in electrical circuits, especially since most X-ray machines are digital today. This is electronic noise. It is often produced by thermal activity in the system and is Gaussian in its distribution \cite{goyal2018}.

As low-dose X-rays have become more and more standard, researchers have increasingly recognized that the actual noise in many X-ray images is resultant of both quantum and electronic noise. This is often modeled as a mixed Poisson-Gaussian noise model \cite{lee2018}.

Three types of noise will be digitally modeled in order to gauge their effect on image classifiers~\cite{boonprong2018classification,gravel2004method,buades2005review}. First, quantum noise, which will be modeled with the Poisson noise filter~\cite{kavuri2020relative,kalivas1999modeling}. Second, electronic noise, which will be modeled by a Gaussian noise filter. And lastly, a combination of the two, which will be modeled by a combination of Poisson and Gaussian noise filters~\cite{iniewski2009medical,goyal2018noise,thanh2015variational,thanh2016method,lanza2014image,le2014unbiased}. These noise models are then progressively added to the input image, and then fed into the model, where several evaluation metrics can inform us on how the performance of each model is changed as the quality of its input is changed. 

\subsection{Baseline Performance}
Under optimal simulated conditions, when neither type of noise is applied to the test data, I found that the fine-tuned pre-trained models demonstrated superior performance over the custom CNN baseline model. This was to be expected, as the custom CNN model is quite simple, with only a handful of layers, and the pre-trained models are huge, with dozens of layers and tens of millions of parameters. 

The fine-tuning process was a great success. Despite small features that can be difficult for even skilled humans to see, each of our three fine-tuned models was able to achieve very high accuracy, AUC, and F1-score. All fine-tuned models achieved initial accuracy scores above 95\%, AUC above 0.99 and F1-scores above 0.95. 

The 'Simple CNN' model also achieved good success on the test dataset. A 92\% accuracy, 0.98 AUC, and an F1-score of 0.93 shows that despite its simple architecture, the model is still highly performance, despite being far less complex than the pre-trained classification models. 

\subsection{Noise Robustness Analysis}

\subsubsection{Gaussian Noise Analysis}
Across all architectures, Gaussian noise proved to be the most disruptive of the two primary types of noise, causing rapid and severe performance degradation. Critical failure thresholds emerged somewhere between the noise levels of 0.001 - 0.0005, depending on the architecture.

EfficientNet and ResNet both failed catastrophically under the progressive Gaussian noise test. ResNet failed almost immediately, even with small amounts of noise. EfficientNet fared slightly better, however, it also failed fast, and did not exhibit a very gradual failure pattern. 

VGG16 exhibited far superior performance under progressive Gaussian noise. It performed far better than both of the other pre-trained models by maintaining reasonable performance metrics for more noise epochs, as well as exhibiting a far more gradual degradation pattern, as we would expect to see. 

The 'Simple CNN' model confirms this expectation. Although it did not outperform the VGG16 model, and was slightly less gradual in its degradation, it outperformed both the ResNet50 model and the EfficientNet model, while also displaying more gradual degradation patterns than them. 

\subsubsection{Poisson Noise Analysis}
All models under consideration exhibited a far greater degree of tolerance to Poisson noise than they did with Gaussian noise. However, the comparative patterns of failure were largely the same.

EfficientNet failed completely. It only took a very small amount of noise to cause total failure, and it failed catastrophically. ResNet failed even earlier, however, it was a bit more gradual in its performance failure. Neither of them performed very well, despite marginal gains when compared to the Gaussian analysis.

VGG16 performed exceptionally well in the Poisson analysis. It maintained a high level of accuracy, AUC and F1 score, all the way up into the very high levels of noise, where the image is nearly destroyed. It remained stable enough for clinical utility for far, far longer than any of the other models. 

The 'Simple CNN' model did demonstrate some consistent gradual degradation and did last far longer than ResNet or EfficientNet, but did not reach the colossal standards set by VGG16 for this analysis.

\subsubsection{Mixed Noise Analysis}
The mixed noise analysis, although it seemed to be the most useful for replicating real world conditions, appeared to produce the least amount of interesting results. The combination of the two noises didn't seem to produce any sort of emergent behavior, as was suggested in other literature. 

As with previous tests, VGG16 emerged victorious. It maintained its long, gradual failing period, while also beating out the other models in terms of performance and how much noise it took to make it fail. 

The 'Simple CNN' model performed slightly better than its Gaussian analysis, but slightly worse than its Poisson analysis. ResNet and EfficientNet performed very poorly with the mixed noise, as with the other two experiments. The VGG16 model was the only one that came away with exceptional or even acceptable results. 

\section{Comparative Model Analysis}
Overall, VGG16 performed the best, according to several metrics. It was by far the most robust against noise, and even when the noise began to break the model down, it still degraded very gradually over time. Looking at Table~\ref{tab:performance_robustness_simple}, we can see that VGG16 was the only pre-trained model which did not catastrophically fail. It's also the only model that still has clinical functionality above 0.001 Gaussian noise. 

The 'Simple CNN' model served its purpose very well. For being as simple as it is, it provided robust performance and degraded gradually, even if it wasn't as gradual as the superior VGG16. It showed good resilience against all types of noise, and failed slower than the ResNet and EfficientNet models. 

Most of the metrics used to keep track of performance followed the accuracy pretty closely. The only small exception to this is the AUC degradation curve, which showed a bit more of a gradual decline than the accuracy. So, our models seem to have retained a bit of discriminative ability, even if the classification accuracy suffered. F1-score followed the accuracy pretty closely, and specifically demonstrated how vulnerable the precision-recall balance was with the more complex models. 

\begin{table*}[h]
\centering
\caption{Performance versus Robustness Summary Across Model Architectures}
\label{tab:performance_robustness_simple}
\begin{tabular}{|>{\raggedright\arraybackslash}p{0.15\linewidth}|>{\centering\arraybackslash}p{0.2\linewidth}|>{\centering\arraybackslash}p{0.2\linewidth}|>{\centering\arraybackslash}p{0.15\linewidth}|>{\centering\arraybackslash}p{0.15\linewidth}|}
\hline
\textbf{Model} & \textbf{Clean Data Performance}& \textbf{Critical Failure Point*}& \textbf{Performance at 0.001 Gaussian}& \textbf{Degradation Pattern}\\
\hline
\textbf{EfficientNet} & 98.62\% (0.9997 AUC) & 0.0001 (75.69\% → 52.77\%) & 47.04\% (Failed) & \textbf{Catastrophic} \\
\hline
\textbf{ResNet50} & 95.06\% (0.9936 AUC) & 5$\times$10$^{-5}$ (88.93\% → 48.81\%) & 47.04\% (Failed) & \textbf{Catastrophic} \\
\hline
\textbf{VGG16} & 95.66\% (0.9921 AUC) & No catastrophic failure & 83.79\% (Functional) & \textbf{Graceful} \\
\hline
\textbf{Simple CNN}& 92.09\% (0.9759 AUC) & No catastrophic failure & 50.00\% (Failed) & \textbf{Graceful} \\
\hline
\end{tabular}
\\[0.5em]
\small{*Critical failure point defined as $>$40\% accuracy drop between consecutive noise levels.}
\end{table*}

Our Simple CNN provides a good baseline that demonstrates what a decently robust architecture might look like when degraded by noise, and the pattern of gradual failing that follows. VGG16 far supersedes "decent" in this case, while ResNet and EfficientNet failed terribly. 

\begin{figure*}
    \centering
    \includegraphics[width=10cm, height=6cm]{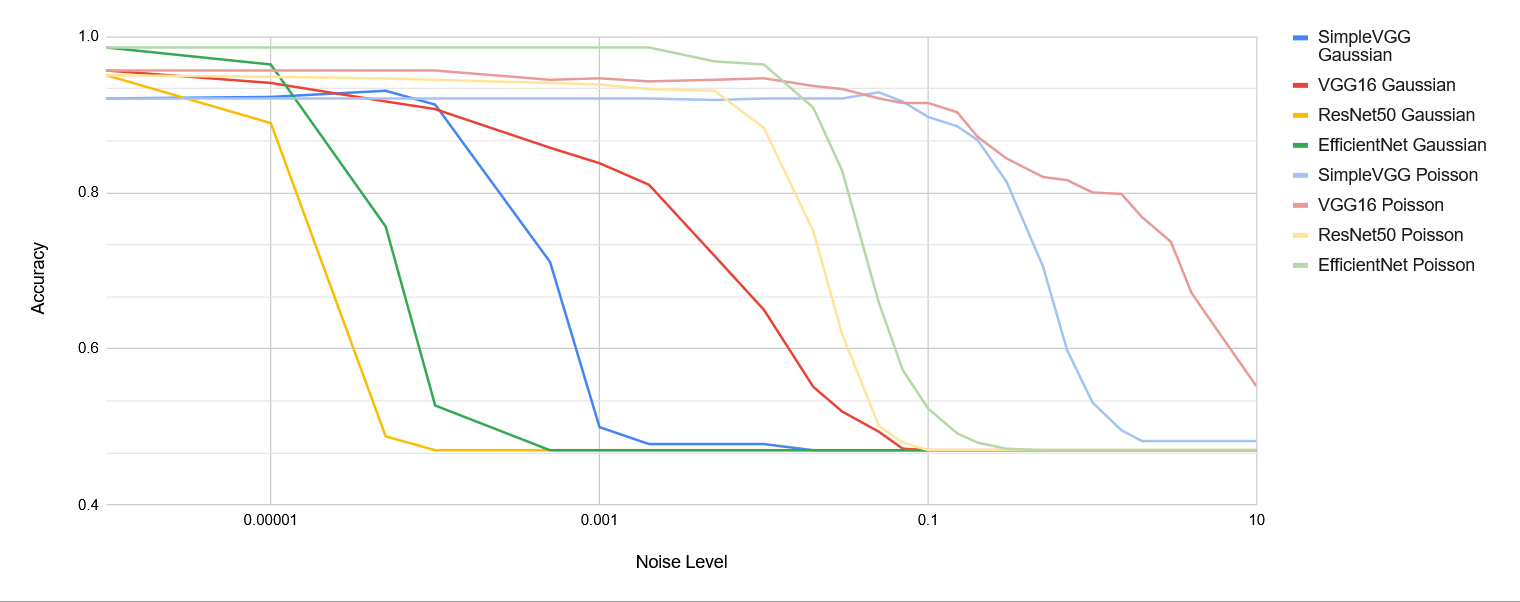}
    \caption{Model accuracy degradation as Gaussian and Poisson noise levels increase over time (log-scale).}
    \label{fig:enter-label5}
\end{figure*}

The results of this thesis reveal an interesting paradoxical tradeoff, which is the tradeoff between complexity and robustness. Looking at Figure~\ref{fig:enter-label5}, we can see that the more sophisticated networks, ResNet and EfficientNet, showed better performance on clean data than the other two networks did, but they were far more brittle under the effects of progressive noise than the simpler two networks. The skip connections of ResNet and the compound scaling techniques of EfficientNet are highly beneficial for clean datasets, which are highly optimized for ImageNet-esque distributions, but this complexity seems to over-fit the feature representation and cause a lack of generalization capability for even slightly degraded inputs. 

Gaussian noise proved to be far more problematic and disruptive in the system. This sort of additive noise may be extra problematic due to the effect it has on the distribution of the image, which many pre-trained models rely heavily on. Poisson noise was far better handled, indicating that the manner in which Poisson noise corrupts the image is far less damaging to the feature space than Gaussian. Mixed noise predominately follows the characteristics of the Gaussian test results, which might suggest that the electronic noise, which the Gaussian noise represents, is a far more limiting factor than quantum noise, despite quantum noise occurring more frequently in the real world. 

Despite its implications on training efficiency, the depth-first approach taken by VGG16  seems to provide more redundant feature pathways that might have been optimized out of systems like EfficientNet. It's also important to note that the skip connections, which are a primary characteristic of ResNet, likely created additional pathways for noise propagation. Overall, the complexity of these networks paradoxically also contributes to their brittleness. 

\section{Healthcare Implications}
A particularly dangerous failure mode was found in the results, which must be addressed. When analyzing the pattern in how precision and recall fail, I have found that complex models often fail by collapsing to zero, rather than having a balanced error increase. This is particularly important to consider in the context of medical imaging, as false negatives can be far more damaging than a false positive. You would generally prefer a few false positives that a doctor could look at and wave away, versus neither the doctor or the software catching it. This is especially important for models trained on healthcare data, which almost always contains more negative examples than positive examples. 

For high resource settings which generally don't have to worry about the quality of medical imaging, a complex solution like EfficientNet would probably be the best solution. However, in resource-limited areas that cannot consistently rely on clear, high-quality medical images, a relatively simpler model like VGG16 should be trained and deployed. It has shown itself here to be far less brittle than more complex architectures due to redundancy inherent to it. This balances the cost efficiency of fine-tuning, while maintaining high levels of performance and reliability. 

\section{Conclusion}
This paper establishes a empirical method of evaluating the robustness of medical AI based in computer vision through a process of progressive noise augmentation which simulates the conditions of real world equipment degradation. A comprehensive assessment of different failure modes is achieved through the study of multiple different types of noise.

Of the architectures studied, this paper finds that VGG16 is the optimal choice of architecture for building a robust bone fracture classification system that can hold up to lower-quality medical imaging. It balances high performance with superior noise tolerance when compared to more complex architectures. It also degrades far more gradually, allowing for more flexibility, where an image with minor noise could still be used. 

This paper highlights the importance of the complexity-robustness trade-off. This when more sophisticated architectures, such as ResNet using skip connections and EfficientNet using compound scaling, achieve higher performance on a clean dataset, but don't generalize as well, especially when confronted with noisy input that reflects the real world conditions in resource limited areas around the globe.

\textit{Study Limitations:} While Gaussian and Poisson noise can capture the most common and most damaging sources of image degradation, there are additional factors not covered by this study, such as structural noise, variations in contrast, blurriness, and geometric distortions. These factors could provide a more advanced insight into the limitations of computer vision machine learning when input can't be guaranteed as consistent or perfect. 

This paper was only performed using the compound dataset mentioned above. These results are specific to this dataset and may not generalize to other forms of medical imaging, or even to a different dataset of X-ray images. This is something to consider, especially with the tendency for these architectures to collapse to zero as they fail. Oversampling the fractured portion of the dataset could address this issue, but further study is needed to determine if that could adversely affect training. 

For the \textit{broader impact}, this paper helps to establish evidence-based guidance for the construction and deployment of applied AI in resource-limited regions, hoping to potentially expand access to diagnostic assistance in these regions who may need it more than anyone. It provides a means of identifying robust architectures to help enable deployment in places where expensive equipment upgrades just aren't viable. 

The identification and emphasis on the tradeoff between complexity and robustness should inform future architectural developments and help emphasize the necessity of robustness alongside accuracy. In some small part, this challenges the zeitgeist, which puts heavy emphasis on optimizing for performance benchmarks, with less regard for real-world implementation.

\bibliographystyle{ieeetr}
\bibliography{fractures_detection}

\end{document}